# An analysis to identify the structural breaks of COVID-19 in Turkey

**Yeşim Güney[1], Yetkin Tuaç[2], Şenay Özdemir[3], Fulya Gökalp Yavuz[4], Olcay Arslan[5]**

ydone@ankara.edu.tr[1], ytuac@ankara.edu.tr[2], senayozdemir@aku.edu.tr[3], fgokalp@metu.edu.tr[4], oarslan@ankara.edu.tr[5],

Ankara University[1,2,5], Afyon Kocatepe University[3], Middle East Technical University[4]

**Abstract**

In countries with a severe outbreak of COVID-19, most governments are considering whether anti-transmission measures are worth social and economic costs. The seriousness of economic costs such as closure of some workplaces, unemployment, reduction in production and social costs such as school closures, disruptions in education could be observable. However, the effect of the measures taken on the spread of the epidemic, such as the number of delayed or prevented cases, could not be observed. For this reason, the direct effects of the measures taken on health, that is, the effects on the course of the epidemic, are important research subjects. For this purpose, in this study, the breakpoint linear regression analysis is performed to analyze the trends of daily active cases, recovered, and deaths in Turkey. The analysis reveals that there has been a remarkable impact on lockdown and other precautions. Using the breakpoint regression model, we also analyze the active cases' trajectory for eight affected countries and compare the patterns in these countries with Turkey.

**Introduction**

COVID-19, the novel coronavirus, was first detected in Wuhan, China, on December 31, 2019, and this disease was characterized as a pandemic on March 11 (WHO, 2020). The new coronavirus infected over 73 million people worldwide, and 1.637.032 people died from the disease by November 25, 2020.

All countries struggling with the COVID-19 outbreak have implemented social distancing measures to curb this epidemic and reduce the mortality rate. The effect of social distancing measures on the spread of the COVID-19 epidemic has become a curious research topic, and the relevant studies quickly took their place in the literature. For example, Siedner et al. (2020) discussed an interrupted time series model to compare the daily growth rate of COVID-19 cases with the post-implementation of social distancing measures in the USA. Weiming et al. (2020) analyzed the changing trends of the COVID-19 outbreak using a joinpoint regression model for Hubei and non-Hubei regions in China. Muggeo et al. (2020) considered segmented regression analysis and wrote R codes to determine the number of breakpoints to measure how much the outbreak's measures slowed the spread of the epidemic. Simsek et al. (2020) analyzed intensive care patient numbers, intubation rates, a positive number of cases and death rates in Turkey between March 27, 2020-April 26, 2020 dates using breakpoint regression and made a comparison to pandemic countries (United States of America, Germany, China, France, South Korea, United Kingdom, Iran, Spain, and Italy) around the world. Santamaria and Hortal (2020) used segmented regression analyses to identify shifts in the evolution of the effective reproduction number reported for 16 Spanish administrative regions and associated these breaking points with a timeline of key containment measures taken by national and regional governments, applying time lags for the time from contagion to case detection, with their associated errors. Senapati et al. (2020) proposed a breakpoint linear regression-based machine learning approach for estimating the number of cases and recovery in five different states in India. Chaurasia (2020) analyzed the trend in daily reported confirmed cases of COVID-19 in India using breakpoint regression analysis. Xie and Zhu (2020) used a piecewise linear regression to determine the relationship between mean temperature and COVID-19 confirmed cases. Additionally, Al Hasan et al. (2020) used breakpoint regression to analyze the temporal trends and identify significant changes in China's COVID-19 outbreak trends. Singh (2020) performed a breakpoint regression analysis in India to understand the pattern of COVID-19.

Most of these studies conducted worldwide have analyzed the early stages of the outbreak, but the outbreak trend has changed several times all around the world. It is now necessary to examine the effects of Turkey's measures and compare the trend of COVID-19 in Turkey with other countries. Therefore, the primary purpose here is to observe the causal impacts of the lockdown imposed in Turkey and eight affected countries around the globe. Therefore, identifying breakpoints is of significant importance. It enables us to characterize the temporal behavior of the COVID-19 and helps policymakers evaluate the effectiveness of the past and ongoing mitigation and intervention strategies. For this purpose, we have considered the breakpoint regression model for active cases, recoveries, and deaths in Turkey. Also, the average daily percent change (ADPC) is calculated as the weighted average of daily percent changes of various segments to quantify the overall changes in active cases. We also compare eight countries with Turkey in terms of the trend of active cases using ADPC.

The rest of the paper is organized as follows: An overview of the breakpoint regression approach is presented in Section 2. Section 3 provides the experimental results, discusses the trend of active cases, recoveries, and deaths in Turkey, and compares the trends of the active cases in Turkey and eight covid-affected countries. Finally, the paper is concluded in Section 4.

**Materials and Methods**

**Data collection**

We focus on Turkey (TUR) and eight countries, including Germany (GER), where the COVID-19 spread is similar to that of Turkey, Iran (IRN), which is a close neighbor of Turkey, China (CHN), the country that the first case was identified in, South Korea (KOR), made good progress fighting off coronavirus, India (IND), Italy (ITA), Russia (RUS) some of the most corona-affected countries, Australia (AUS) almost eliminated the coronavirus by strict measures. The data sets are taken from the COVID19 R package (Guidotti and Ardia, 2020), where cumulative measures such as confirmed cases, recoveries, and deaths are updated daily for each nation. For analysis part of this study, the number of daily active cases, recoveries, and deaths are considered from the date which the number of cases reached 100 to November 25, 2020. The number of daily cases, recovered, deaths was obtained by taking the difference between the new day total reported numbers and the total reported numbers from the previous day. The total number of active cases equal to the total number of confirmed cases minus the total number of deaths minus the total number of recovered. The annotated R codes for this analysis are available at http://covid19stats.ankara.edu.tr/ .

**Statistical Modeling**

Breakpoint regression is a useful tool for assessing the effect of a covariate x on the response y when x changes at a value within the range of the covariate. More generally, segmented modeling represents a useful framework in many areas, including climatology, environmental epidemiology, and ecology (see Muggeo, 2017 and references therein). The breakpoint regression model has also been widely used in medical and related studies such as mortality (Parcha et al., 2020), infections (Smiddy et al., 2020), cancer studies (Mesquita et al., 2020), medication usage (Wagner et al., 2002), etc. Therefore, we consider the breakpoint regression model to analyze the trend of the COVID-19 in Turkey and some other countries that have been severely affected by COVID-19 pandemic.

The breakpoint regression model for the sample $(x_t, y_t, t = 1,2,\ldots,n)$ is defined as follows.

$$y_t = \begin{cases} m\left(x_t, \underline{\beta}^{(1)}\right) , & a \leq x_t \leq r \\ m\left(x_t, \underline{\beta}^{(2)}\right) , & r < x_t \leq b \end{cases} \quad (1)$$

where $\underline{\beta}^{(1)} = \left(\beta_0^{(1)}, \beta_1^{(1)}\right)'$ ve $\underline{\beta}^{(2)} = \left(\beta_0^{(2)}, \beta_1^{(2)}\right)'$.

We consider $m(\cdot)$ as a linear function of $x_t$, i.e., $m\left(x_t, \underline{\beta}\right) = \beta_0 + \beta_1 x_t$. Here, $\beta_1$ is the slope of the regression line, which is an important parameter that determines the increasing or decreasing trend of the disease for each time interval. In more than one breakpoint, the breakpoint linear regression model is defined as follows.

$$y_t = \beta_0 + \beta_1 x_t + \delta_1 (x_t - \tau_1)^+ + \cdots + \delta_k (x_t - \tau_k)^+, t = 1,2,\ldots,n \quad (2)$$

In this study, we consider $x_t$ as the number of the day. Here, $\tau_s$'s $(s = 1,2,\ldots,k)$ are the breakpoints. In this case, k breakpoints divide the time into the $(k + 1)$ interval, and the slopes associated with these intervals are $\beta_1$ (first slope, $x_t < \tau_1$)), $\beta_2 = \beta_1 + \delta_1$ (second slope), and so on. The final slope is defined as $\beta_{k+1} = \beta_1 + \delta_1 + \cdots + \delta_k$.

All statistical analyses were performed using R version 4.0.3 (R Core Team, 2020). Parameter estimates and breakpoint estimates were calculated with "segmented" package (Muggeo, 2008). Initial values for breakpoints were calculated with "strucchange" package (Zeileis et al., 2002). We first determine the data's change points using the "strucchange" package (Zeileis et al., 2002), then we use these points as initial breakpoints in the "segmented" package in R.

**Results**

**Analysis of Turkey COVID-19 Data**

This section discusses breakpoint regression analysis for daily active cases, recoveries, and deaths in Turkey. Firstly, we give the timeline of important events and measures during the COVID-19 pandemic in Turkey in Figure 1.

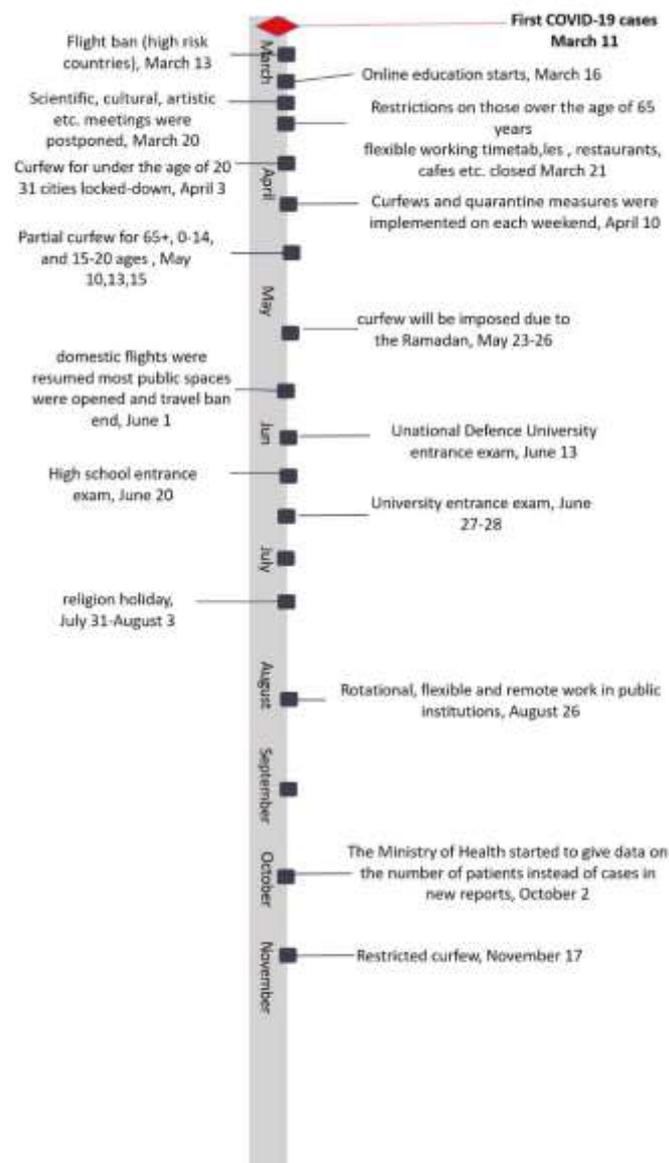

Figure 1. Timeline of the key events for the spread of COVID-19 in Turkey

Figure 1 shows the important events that affect the spread of COVID-19 in Turkey and offers the precautions against to COVID-19. Figure 1 is created with the timeline given in https://en.wikipedia.org/wiki/Timeline_of_the_COVID-19_pandemic_in_Turkey.

This timeline will be used to understand which events or restrictions are related to the breakpoints of the number of daily active cases, recoveries, and deaths in Turkey. By doing so, new policies can be developed against the epidemic, and a new timeline can be created for these policies. For this reason, we analyze the number of daily active cases, recoveries, and deaths in Turkey using a breakpoint regression model. Turkey COVID-19 data was considered since March 18, 2020, when the total number of cases reached 98 until November 25, 2020.

**Active cases.** For the number of active cases in Turkey, parameter estimates, estimates of breakpoints, and the date corresponding to the breakpoints' estimates are given in Table 1.

Table 1. Breakpoint regression estimates and dates for the active cases in Turkey

| Parameter estimates | Breakpoint estimates | Dates |
|---|---|---|
| 2523.0 | 38 | 24.04.2020 |
| -2632.0 | 55 | 11.05.2020 |
| -359.2 | 136 | 31.07.2020 |
| 360.5 | 244 | 15.11.2020 |
| 2218.0 | | |

In Figure 2, the number of active cases in Turkey, fitted regression lines, breaking points, and confidence intervals related to the breaking points (on top) are given.

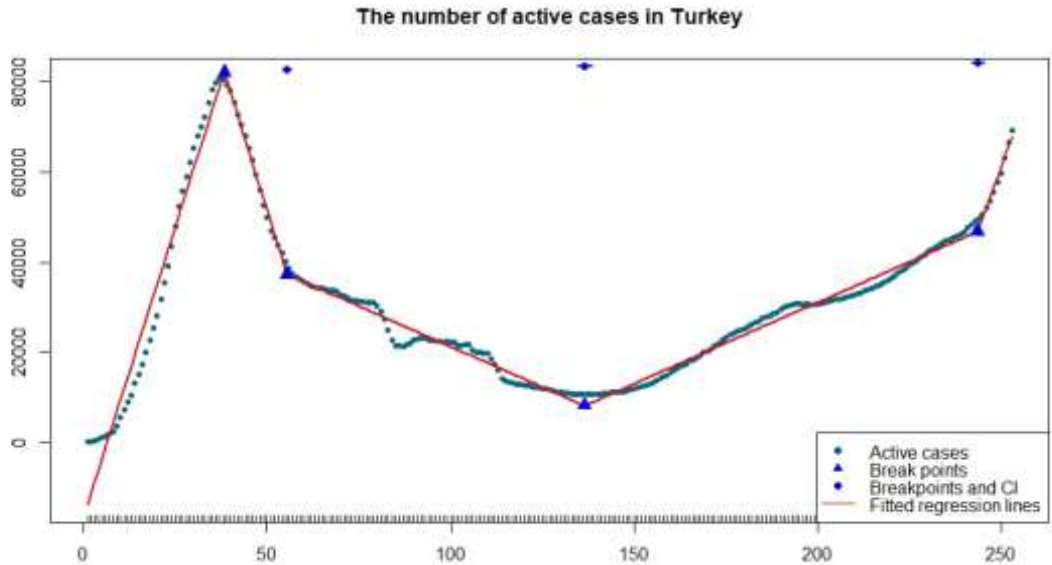

Figure 2. Fitted breakpoint regression lines for the number of active cases in Turkey

Considering the estimated dates given in Table 2, it can be said that there is a rapid increase in the number of active cases from this date until April 24, 2020 (slope = 2573). It is seen that a sharp decrease started approximately as of April 24, 2020 (slope = -2632). This sharp decrease is due to the decision to suspend face-to-face education on March 23, 2020, the partial curfew on April 3, 2020, the general curfew on April 11-12, 2020, and the following official holidays and religious holidays. It can be interpreted as the effect of curfews. As seen from the regression graph, the decrease in the number of active cases continue rapidly between April 24, 2020, and May 11, 2020. The decrease in the number of active cases slows down from May 11, 2020, to July 31, 2020 (slope = -359). With the normalization decisions taken on May 4, 2020, it can be interpreted that the speed of this decrease has decreased as a result of the easing of some measures. After July 31, 2020, a sharp increase starts (slope = 360), and this increase accelerate as of November 15, 2020 (slope = 2218). The increase in the number of active cases, which started on July 31, 2020, can be interpreted as the complete abolition of the measures, the release of intercity travel, and the release of domestic and international holiday travels. The rapid increase after November 15, 2020, due to the change in season, people are mostly in closed areas and do not take other serious measures other than masks, distance, and hygiene measures.

**Daily Recovered**. The number of daily recoveries was also analyzed using the breakpoint linear regression model. The parameter estimates, estimates of breakpoints, and the date corresponding to the breakpoints' estimates are given in Table 2.

Table 2. Breakpoint regression estimates and dates for the daily recoveries in Turkey

| Parameter estimates | Breakpoint estimates | Dates |
|---|---|---|
| 119.700 | 49 | 5.05.2020 |
| -255.000 | 60 | 16.05.2020 |
| -6.435 | 162 | 26.08.2020 |
| 9.774 | 233 | 5.11.2020 |
| 108.000 | | |

In Figure 3, the number of daily recovered numbers in Turkey, fitted regression lines, breaking points, and confidence intervals related to the breaking points (on top) are given.

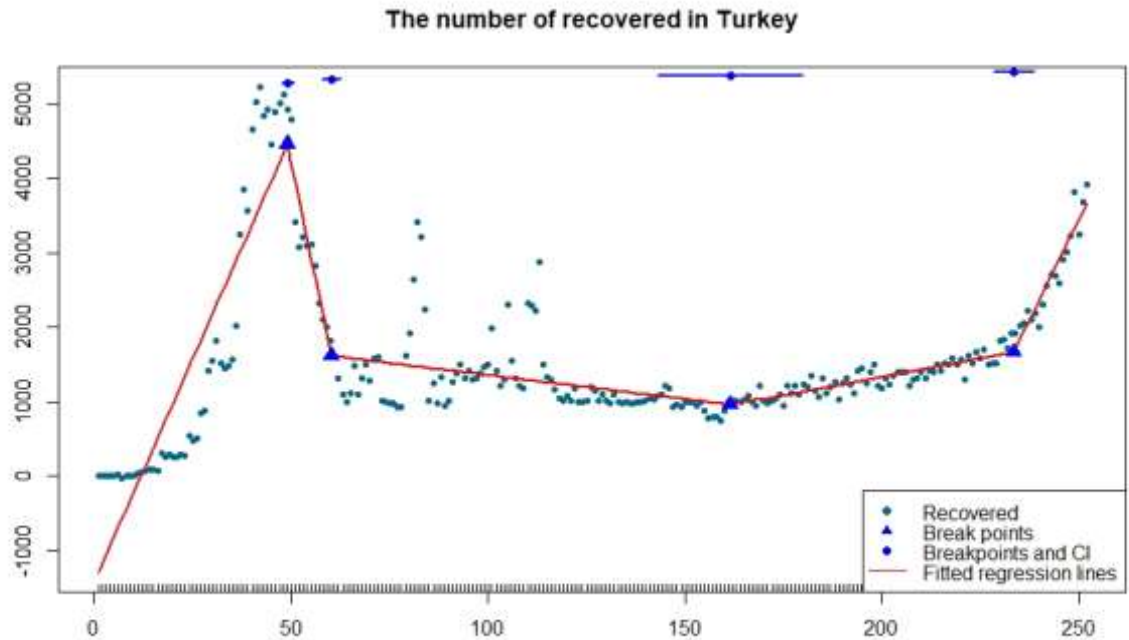

Figure 3. Fitted breakpoint regression lines for the number of daily recoveries in Turkey

Considering the estimated dates given in Table 3, the number of daily recoveries increased rapidly from March 18, 2020, to May 5, 2020 (slope = 119.7). Between May 5, 2020, and May 16, 2020, there is a sharp decrease in daily recoveries (slope = -255). The reason behind this decrease can be the decrease in the number of active cases as of April 24, 2020 (see Table 1). This interpretation is based on the assumption that the recovery time of the disease is approximately two weeks. It is an expected result that the number of recoveries will decrease around two weeks after April 24, 2020 (the approximate recovery period of the disease), that is, as of May 5, 2020, when active cases started to decrease. This decrease in daily recoveries continued slowly from May 16 16, 2020, until August 26, 2020 (slope = -6.435). Since the date August 26, 2020, the number of recoveries increase (slope = 9.774), and then after November 11, 2020, this increase was significantly accelerated (slope = 108).

**Daily deaths.** The number of daily deaths was analyzed by the breakpoint linear regression model. The parameter estimates, estimates of breakpoints, and the date corresponding to the breakpoints' estimates are given in Table 3.

Table 3. Breakpoint regression estimates and dates for the daily deaths in Turkey

| Parameter estimates | Breakpoint estimates | Dates |
|:---:|:---:|:---:|
| 4.587 | 30 | 16.04.2020 |
| -3.114 | 63 | 19.05.2020 |
| -0.145 | 133 | 28.07.2020 |
| 0.696 | 242 | 14.11.2020 |
| 7.873 | | |

The number of daily deaths in Turkey, fitted regression lines, breaking points, and confidence intervals related to the breaking points (on top) are given in Figure 4.

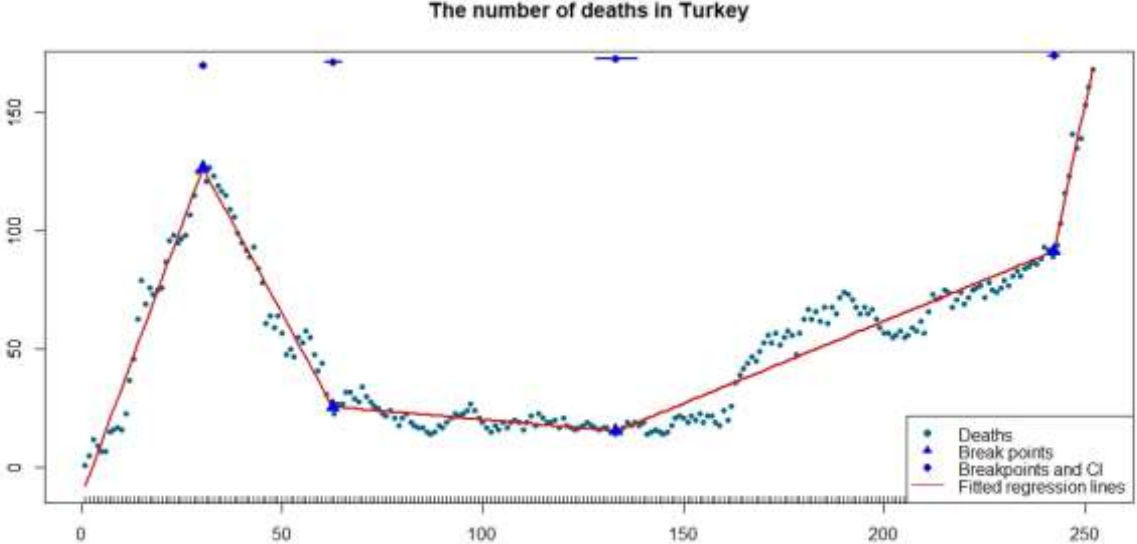

Figure 4. Fitted breakpoint regression lines for the number of daily deaths in Turkey

Considering the estimated dates given in Table 4, the number of daily deaths has increased from March 18, 2020, to April 16, 2020 (slope = 4.587). The number of daily deaths decreased sharply until May 19, 2020 (slope = -3.114). Between May 19, 2020, and July 28, 2020, the decrease in daily deaths slowed down (slope = -0.145). As an estimate, the number of deaths started to increase again since July 28, 2020 (slope = 0.696), and this increase accelerated significantly since November 14, 2020.

**Comparison of Turkey and Other COVID-19-affected Countries**

As point out in the study of Chaurasia (2020), this method has generally been applied with the calendar year as the time scale in these studies. One way to characterize the trend over time is APC (annual percent change). However, since the COVID-19 data are available daily, not yearly, the APC will be DPC (daily percent change) in this context (Chaurasia, 2020). The DPC from day to day (t+1) is defined as follows.

$$\text{DPC} = \left(\frac{y_{t+1} - y_t}{y_t}\right) \times 100 \quad (3)$$

If the trend in the active cases of COVID-19 is modeled as

$$\ln(y_t) = \beta_0 + \beta_1 x_t, \quad t = 1, 2, \ldots, n \quad (4)$$

Then DPC will be as follows.

$$\text{DPC} = (e^{\beta_1} - 1) \times 100 \quad (5)$$

To summarize the fitted regression relationships, Clegg et al. (2009) proposed "average annual percent change (AAPC)" as a scalar calculation of change. In this case, the AAPC will be the average daily percent change (ADPC), which is calculated using a weighted average of DPC in different time segments of the reference period with the weights equal to the length of each segment over the interval (Chaurasia, 2020). DPC and ADPC provide summary statistics of the trend. The DPC value of r% means that the case rates change at r% of the rate per day. The ADPC provides a summary measure of the DPCs over a period where the trend is not constant.

In this part, we compare the trends of the number of active cases per 1 million population of Turkey and eight other countries using ADPC. In this part, for each country, we consider the log-linear model given in equation (4). For each country, COVID-19 data was considered since the total number of cases reached approximately 100 until November 25, 2020. The estimated ADPC values, standard errors, lower and upper bounds of 95 percent confidence intervals for ADPCs are reported in Table 4. Figures 5-12 show the logarithm of the number of active cases per 1 million, fitted regression lines, breaking points, and confidence intervals related to the breaking points (on top) for each country are given in the Appendix.

Table 4. The estimated ADPCs values for the active cases per 1 million population

| Country | ADPC (%) | SE | CI-L | CI-U |
|---------|----------|----|------|------|
| AUS | 1.128 | 0.02721 | 1.074 | 1.181 |
| CHN | 0.1501 | 0.03192 | 0.08757 | 0.2127 |
| GER | 3.905 | 0.05005 | 3.807 | 4.003 |
| IND | 3.306 | 0.009359 | 3.288 | 3.325 |
| IRN | 2.918 | 0.02171 | 2.876 | 2.961 |
| ITA | 2.969 | 0.02078 | 2.928 | 3.009 |
| KOR | 1.171 | 0.02731 | 1.117 | 1.224 |
| RUS | 3.318 | 0.01326 | 3.292 | 3.344 |
| TUR | 2.113 | 0.03411 | 2.046 | 2.18 |

The ADPCs given in Table 4 are positive, which means that the number of active cases in each country has an increasing trend. The joinpoint regression analysis suggests that the logarithm of Turkey's active cases increased at an ADPC of slightly more than 2 percent during Mach 18, 2020-November 25, 2020. According to ADPCs given in Table 4, the daily increase for the active cases was the highest for Germany and the lowest for China for the days between the day reached 100 cases and November 25, 2020. The increase in new cases in Australia is similar to that in South Korea, but it is significantly lower than that in other countries except for China. The ADPC value of Turkey is closest to that of Iran.

**Conclusion and Discussion**

Keeping under control of the spread of COVID-19 requires the timely and effective assessment of the precautions. In all countries affected by COVID-19, measures, such as quarantine, social distancing, and locked-down policies have been executed at various phases to prevent the virus's transmission. It is essential to determine the effects of not following these new policies against the virus. For this reason, we propose to use the breakpoint regression model to characterize the trend of the trajectory of the active cases, recoveries, and deaths for Turkey. We also perform the analysis for eight affected countries and compare the patterns in these countries with Turkey through the estimated ADPCs.

According to the results, the breakpoint regression model's flexibility could be beneficial in adequately describing the active cases in Turkey due to the COVID-19. We found that implementation of social distancing measures was associated with a reduction in the number of active cases between April 24, 2020, and May 11, 2020. This decrease is due to the decision to suspend face-to-face education, the partial curfew, the general curfew on April 11-12, 2020, and the curfew on the following official holidays and religious holidays. It can be seen from the timeline given in Figure 1. In addition, it may be said that because of the complete abolition of the measures, the release of intercity travel, the release of domestic and international holiday travels, an increase starts in the number of active cases in Turkey after July 31, 2020.

We also observed that the breakpoint regression model might provide a good fit to the active case data set for each considered country, and it is useful for early alerts against battling COVID-19. According to the ADPCs given in Table 4, the number of active cases in each country has increased trend, it can be suggested that hard measures should be taken to slow down the spread of the transmission of the virus.

**Acknowledgment**

This study is funded by TUBITAK (The Scientific and Technological Research Council of Turkey) through Project no: 120K594.

**Appendix**

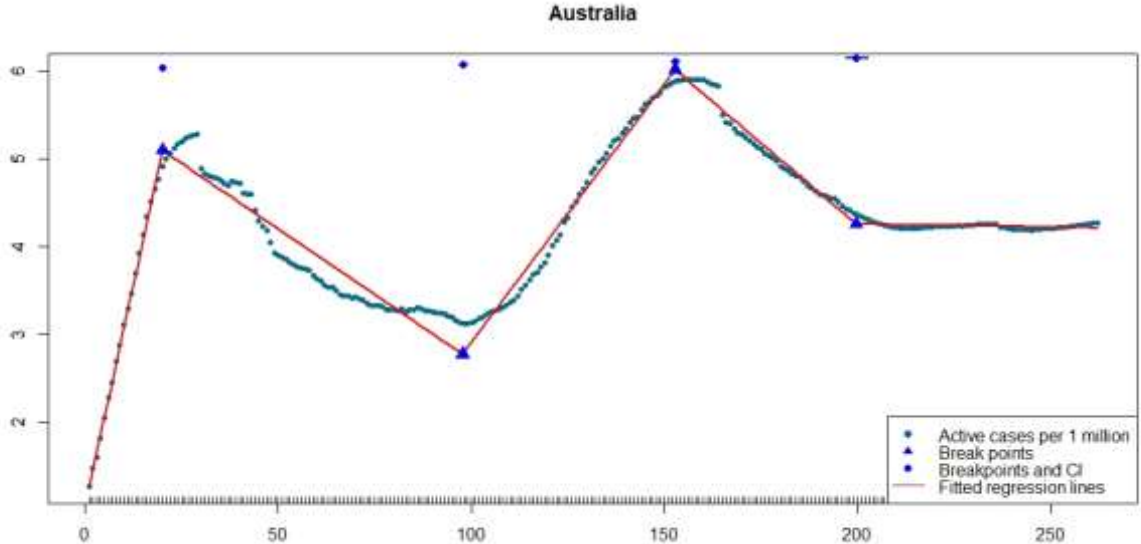

Figure 5. Fitted breakpoint regression lines for the logarithm of the number of active cases per 1 million population in Australia

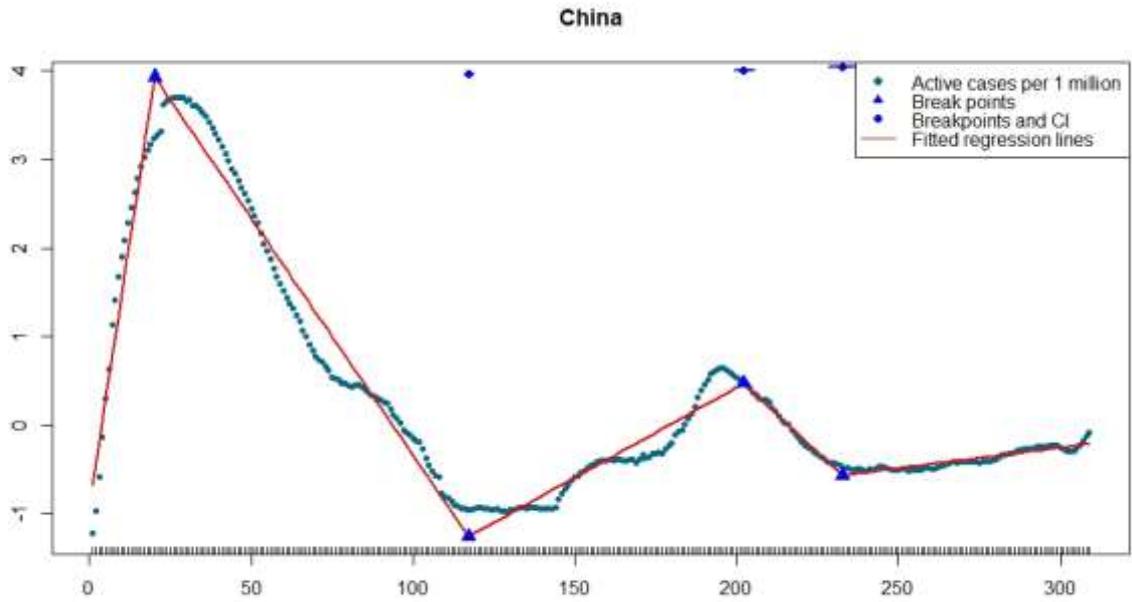

Figure 6. Fitted breakpoint regression lines for the logarithm of the number of active cases per 1 million population in China

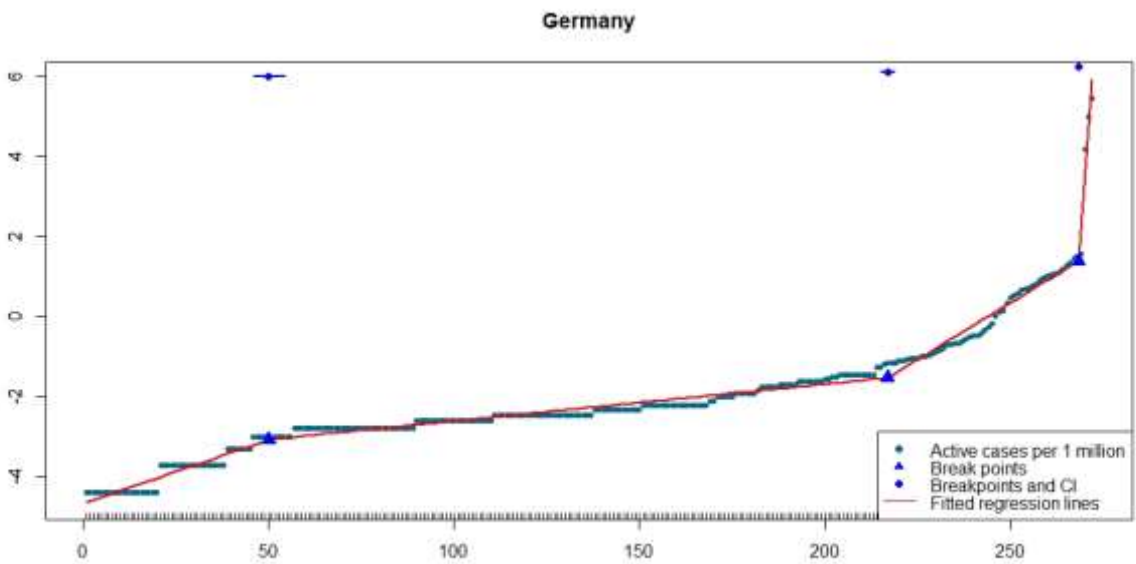

Figure 7. Fitted breakpoint regression lines for the logarithm of the number of active cases per 1 million population in Germany

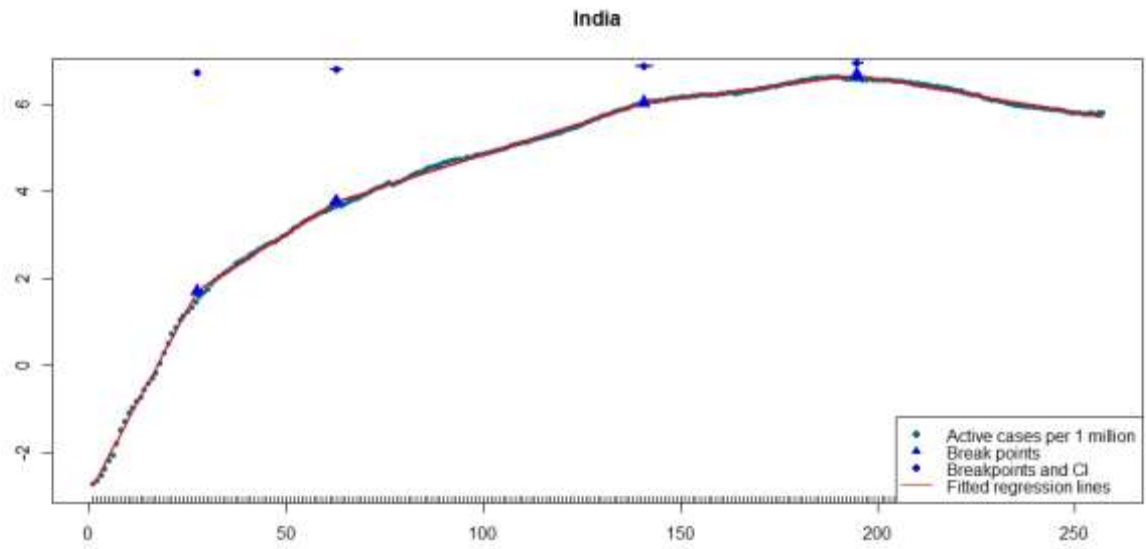

Figure 8. Fitted breakpoint regression lines for the logarithm of the number of active cases per 1 million population in India

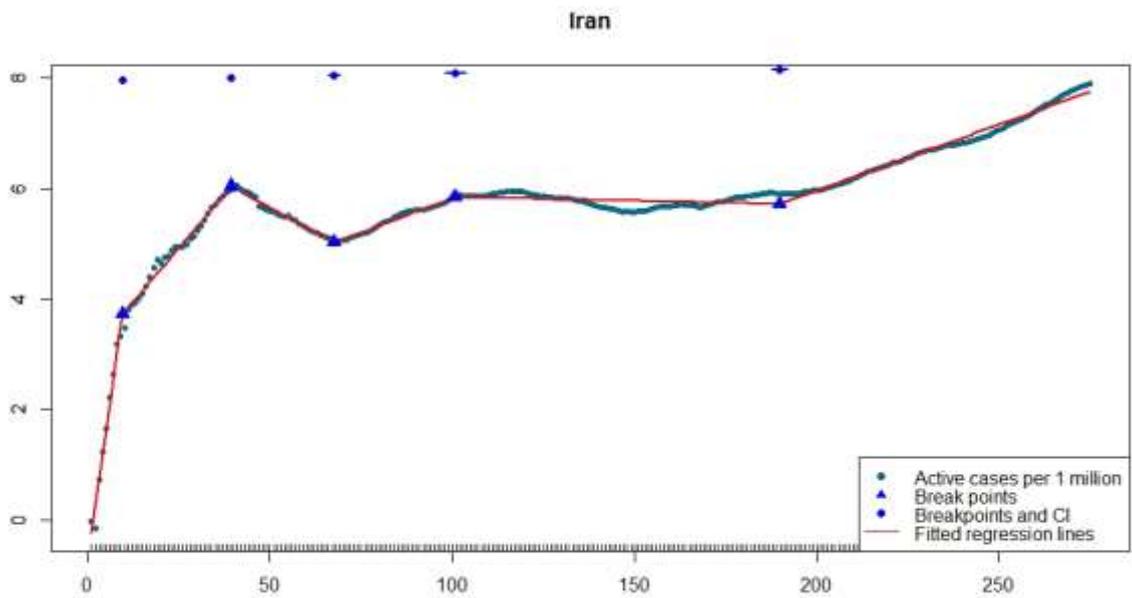

Figure 9. Fitted breakpoint regression lines for the logarithm of the number of active cases per 1 million population in Iran

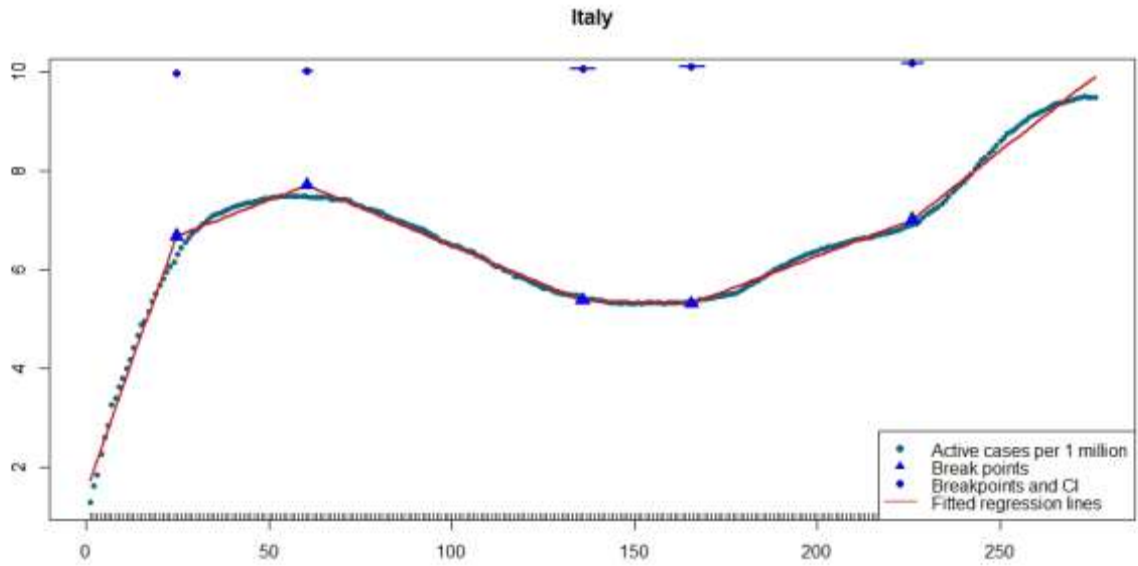

Figure 10. Fitted breakpoint regression lines for the logarithm of the number of active cases per 1 million population in Italy

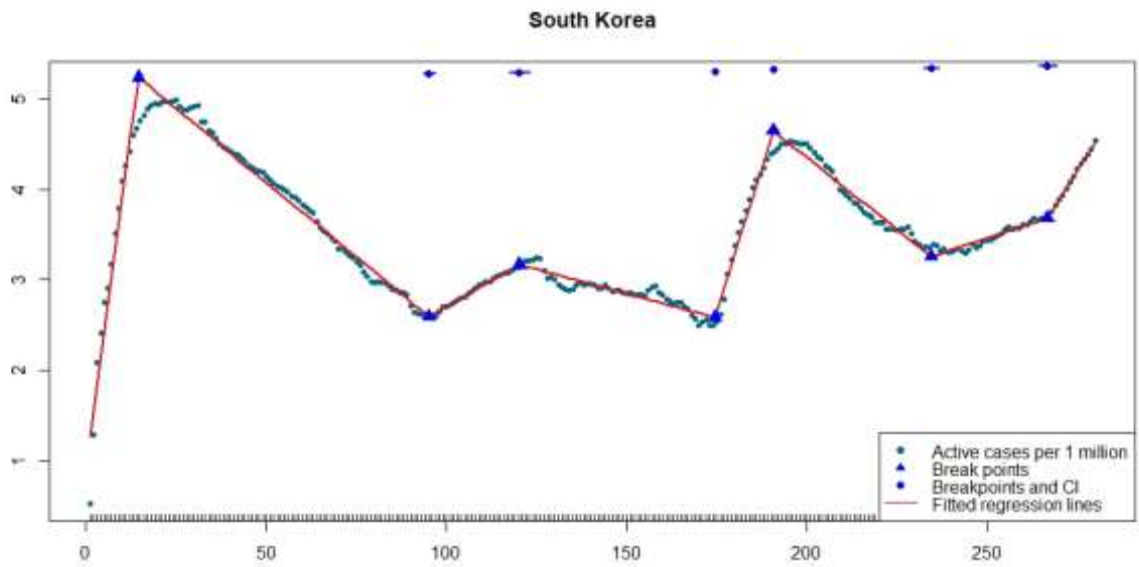

Figure 11. Fitted breakpoint regression lines for the logarithm of the number of active cases per 1 million population in South Korea

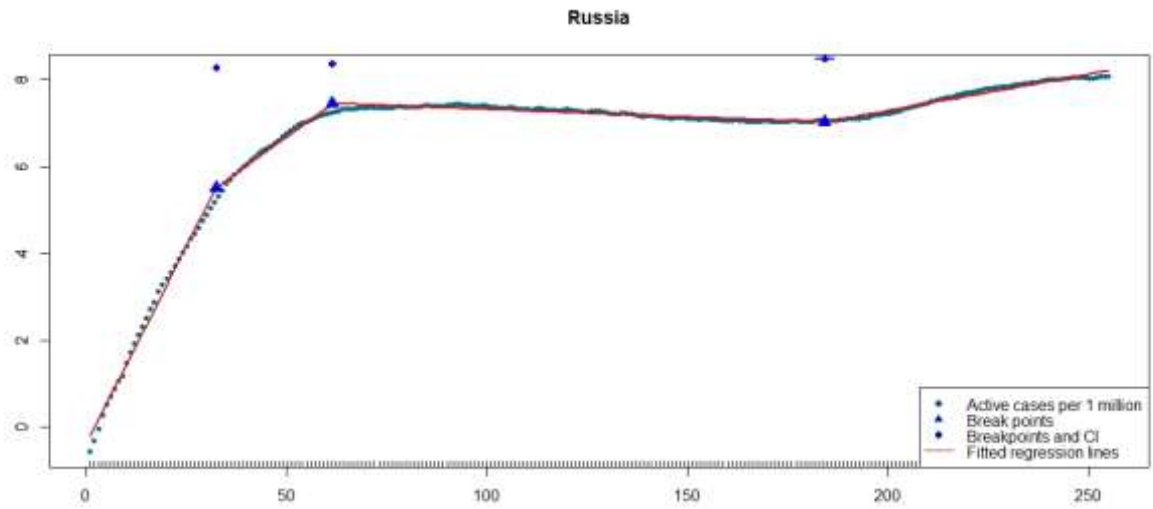

Figure 12. Fitted breakpoint regression lines for the logarithm of the number of active cases per 1 million population in Russia